\documentclass[12pt]{iopart}

\usepackage{graphics}
\usepackage[pdftex]{graphicx}
\usepackage{epstopdf}
\begin{document}

\title{Experimental Observation of a Self-Heating Effect of $^4$He Superflow}

\author{Yongle Yu \dag\  and Hailin Luo \ddag}
\address{\dag\ State Key Laboratory of Magnetic Resonance and Atomic and \\
Molecular Physics,
  Wuhan Institute of Physics and Mathematics, \\Chinese Academy of Science,
 West No. 30 Xiao Hong Shan, Wuchang, \\ Wuhan, 430071, China}
\address{\ddag\ Fujian Institute of Research on the Structure of 
 Matter,\\ Chinese Academy of Sciences, 155 Yangqiao Road West, Fuzhou, 350002, China}

\begin{abstract}
  we report  a counter-intuitive  self-heating effect of
 $^4$He superflow.
This fundamentally unusual heating effect bears a phenomenological resemblance 
to the Peltier
effect of electric current across two different conductors. 
It reveals that 
$^4$He superflow carries thermal energy and entropy,
which is in contrast to the two-fluid model of superfluid $^4$He. 
A natural understanding of this heating effect is
provided by a recently developed microscopic quantum
theory of superfluid $^4$He.

\end{abstract}
\pacs{ 05.70.-a, 67.25.dg, 67.40.Kh}
\vspace{2pc}
\noindent{\it Keywords}: superflow, entropy, quantum many-body physics\\
\vspace{2pc}

\vspace{2pc}

As one of the most fundamental quantum systems in condensed matter 
physics, superfluid $^4$He \cite{kapitza, allen} exhibits a wide variety of extraordinary 
behaviors.  Some intriguing phenomena of superfluid $^4$He, such 
as the fountain effect \cite{allenfountain} and the mechano-colaric effect \cite{mendelssohn}, 
reveal a fundamental
coupling between thermodynamic motion 
and the hydrodynamic motion of the system. Pursuing a quantum 
understanding of this unusual 
coupling is
challenging but nevertheless important. In the past, people 
rely on the two-fluid model  \cite{tisza, landau} of superfluid $^4$He 
to describe physics raised by this coupling. In spite of 
its phenomenological success and its capability to explain
many interesting behaviors of superfluid $^4$He observed later (see {\it{e.g.}}
\cite{diribaene, midlik, varga, vidal, jelatis, beamish, chan, hallock,reppy}), 
 the two-fluid model is not completely 
flawless by itself. It involves a hypothesis of
a macroscopic sub-system ({\it{i.e.}} superfluid component)
 with zero entropy, which is highly exotic. The zero-entropy 
hypothesis implies unavoidably that
this sub-system has a temperature of absolute zero, but
it is difficult to imagine  how  such a sub-system 
can coexist with its thermal surroundings.  Some 
experiments \cite{osborne, androni} questioned
 the validity of the two-fluid model in the past.
 In this work, we report a counter-intuitive self-heating effect
 of $^4$He superflow. 
This fundamental effect reveals that  $^4$He superflow carries 
thermal energy and entropy, 
which is in contrast to the hypothesis of the two-fluid model. 


The main setup of the superflow system is schematically plotted 
in Fig. \ref{fig:schematic}. Three vessels (referred to as pot $A$, 
pot $B$ and cell $C$) are 
connected in series by two superleaks (referred to as $S_{AC}$ and $S_{BC}$).
 Cell $C$ is thermally isolated 
from its surroundings, except its thermal links to the pots via the 
superleaks. In the experiment, pot $A$ is filled with superfluid $^4$He 
initially, then 
superflows through $S_{AC}$, cell $C$ and $S_{BC}$ can be established 
eventually by setting a positive 
temperature difference between pot $B$ and $A$ ({\it{i.e.}} fountain effect), 
resulting in a superfluid transport
from pot $A$ to $B$. If superflow carries zero thermal energy,  
the temperature of cell $C$ ($T_C$) shall eventually lie between the temperature 
of pot $B$ 
($T_B$) and that of pot $A$ ($T_A$). However, it is observed that cell $C$ can be
 heated strikingly by the superflows; as a result, $T_C$ reaches a steady value 
which exceeds $T_B$ by more than one hundred millikelvins.

\begin{figure}%
\begin{center}
\includegraphics[scale=0.2]{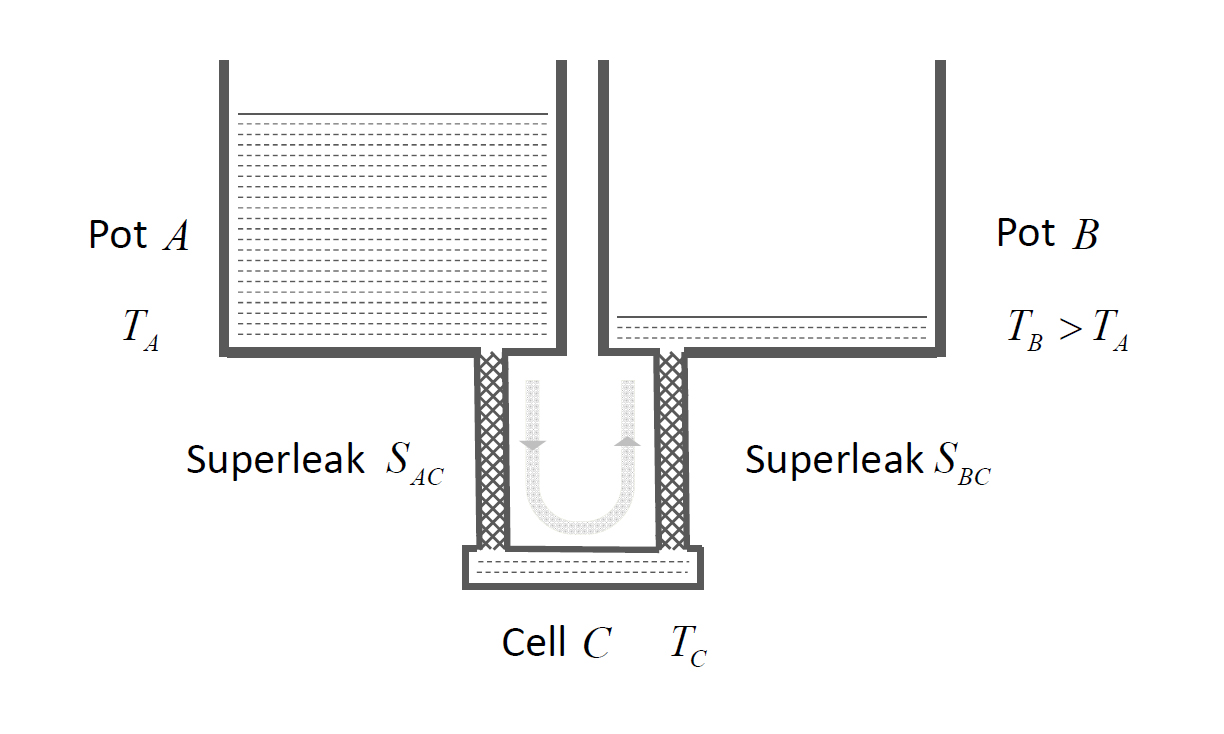}%
\caption{A schematic plot of superflow system.
} %
\label{fig:sflevels}%
\end{center}
\end{figure}

\begin{figure}%
\begin{center}
\includegraphics[scale=0.15]{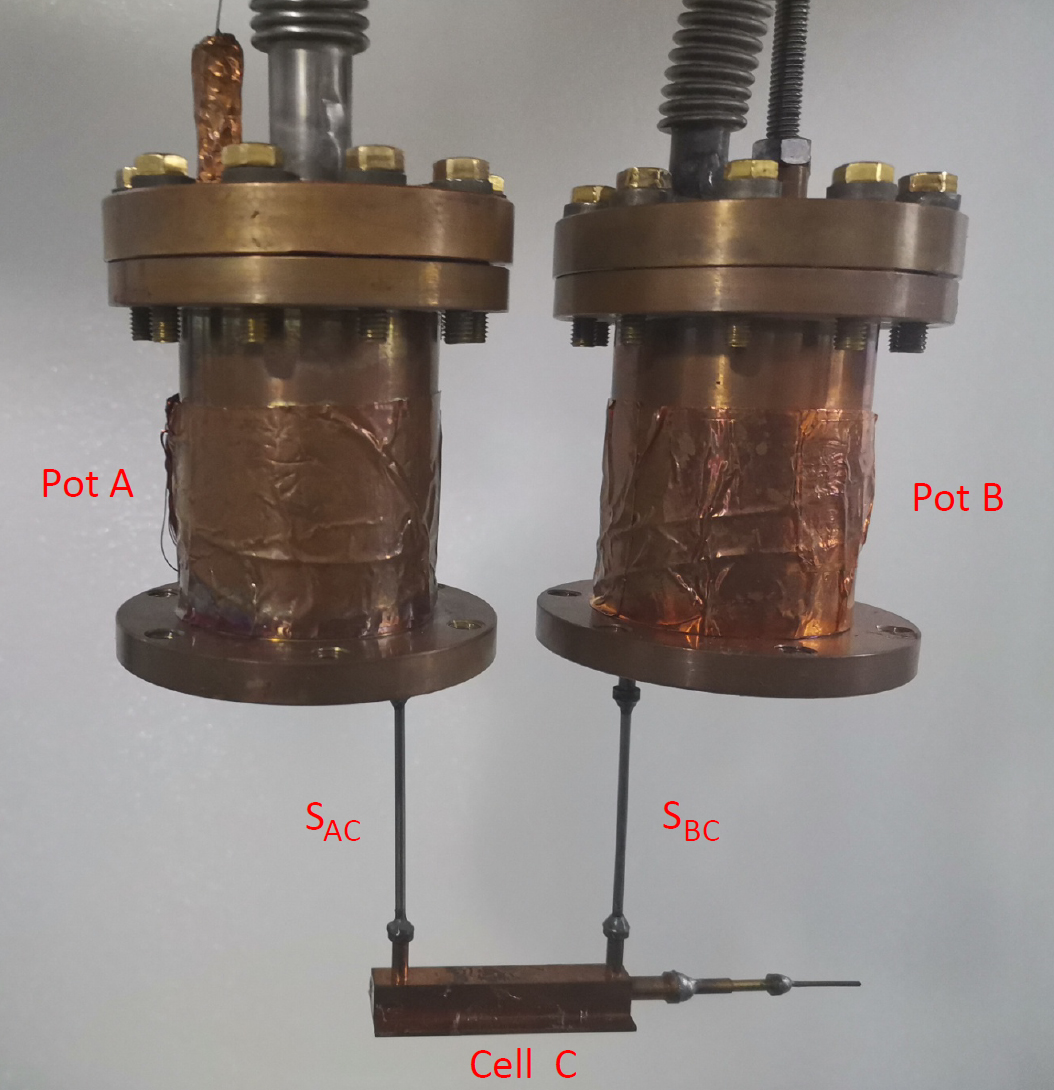}
\caption{A picture of superflow system. $S_{AC}$ and $S_{BC}$ refer to the superleaks.
} %
\label{fig:photos}%
\end{center}
\end{figure}

The experiment is carried out on a  two-stage Gifford-McMahon 
refrigerator with a cooling power of 1 $W$ at 4.2 $K$ and a base temperature of 2.4 $K$. In
order to reach the superfluid temperature regime, a liquid $^4$He cryostat is constructed 
by following largely the design given in  Ref. \cite{field}.  A stainless 
steel capillary, with 
an inner diameter (i.d.) of 0.18 $mm$, an outer diameter (o.d.) of 0.4 $mm$ and a length
 of 1 $m$, is used as the Joule-Thomson 
impedance in the cryostat. The copper pot for collecting liquid $^4$He, with an i.d. 
of 4.0 $cm$ and a volume of 78 $cm^{3}$, is also served as pot $A$ for the experiment. Another 
copper pot, identical to pot $A$, is used as pot $B$. Cell $C$ is made of a small 
copper block, and the main part of its inner cavity is cylindrical, with 
a diameter of 3 $mm$ and a length of 40 $mm$. Each superleak is made of a stainless 
steel tube packed with jeweler's 
rouge powder (with an average particle size of 70 $nm$ determined by TEM).
The tube for   $S_{AC}$ has an i.d. of 0.8 $mm$, an o.d. of 2.0 $mm$
and a length of 65 $mm$, while the tube for $S_{BC}$ has an i.d. of 1.0 $mm$, 
an o.d. of 2.0 $mm$
and a length of 65 $mm$.
 Two superleaks are soft soldered to cell $C$, and they are positioned in
 a way so that the lower end of each superleak  is in proximity to one
end of cell $C$'s cylindrical cavity (see Fig. \ref{fig:photos}). The upper end of
  $S_{AC}$ joins pot $A$ while the upper end of $S_{BC}$ 
joins pot $B$.

A combination of copper braids and brass
strips is used as a thermal link between pot $A$ and a cooling plate directly
 mounted to the second stage of the refrigerator, with a thermal conductance
 of around 2 $mW/K$ at 2 $K$. Pot $B$'s major thermal link with its surroundings 
is a copper braid joining two pots at ends. Resistance wires wrapped around pots are
 used as heaters.  Pot $B$ as well as pot $A$ is equipped with a pumping line.
  Valves are used in the lines so that the pumping rate can be 
manipulated to provide further means for the temperature controls of the
pots. Calibrated carbon ceramic resistances \cite{ccs} are used as temperature 
sensors  to measure $T_A$,  $T_B$ and $T_C$, 
with an accuracy of 5 $mK$. The 
dissipation power of temperature senor on cell $C$ is kept well below $10^{-7}$ 
 $W$,
 so that its heating effect is very limited. 

Two radiation shields made of copper coated with nickel thin layer are amounted
to two stages of the refrigerator, with one at a temperature around $45K$ and the other
at a temperature below $2.8K$. Great care is made to prevent cell $C$ from 
exposing to the thermal radiations from sources at  temperature above $3K$.
Comparative experiments were conducted  to rule out that
background thermal radiation is responsible for the heating
phenomenon observed.

For the initial accumulation of liquid $^4$He (with a purity of 99.999\%) in pot $A$, 
 $T_A$
is raised above $\lambda$ point to  prevent  superflow    
through $S_{AC}$.  
After decreasing $T_A$ below
  $\lambda$ point, superflows through superleaks and cell $C$ can be established by 
setting a large temperature difference
 between $T_B$ and $T_A$.
At given $T_A$ and $T_B$, superflows are let to flow for long enough without
disruption so that $T_C$ can reach its steady value .
Some steady 
values of $T_C$ are listed in Tab. \ref{tab:Ts}. 

\begin{table}
\centering
\begin{tabular}[c]{|l|c|r|}
\hline 
$T_A$ ($K$) & $T_B$ ($K$) & $T_C$ ($K$) \\ \hline 
1.500(4) &  1.700(4)  & 1.847 (1) \\ \hline
1.600(4) &  1.800(4)  & 1.927 (1)\\ \hline
1.600(4) & 1.900(4)   & 2.014 (1)\\ \hline
\end{tabular}
\caption{Steady values of $T_C$ at given $T_A$ and $T_B$. }
\label{tab:Ts}
\end{table}
 
\begin{figure}[htbp]
\begin{center}
\includegraphics[scale=1.0]{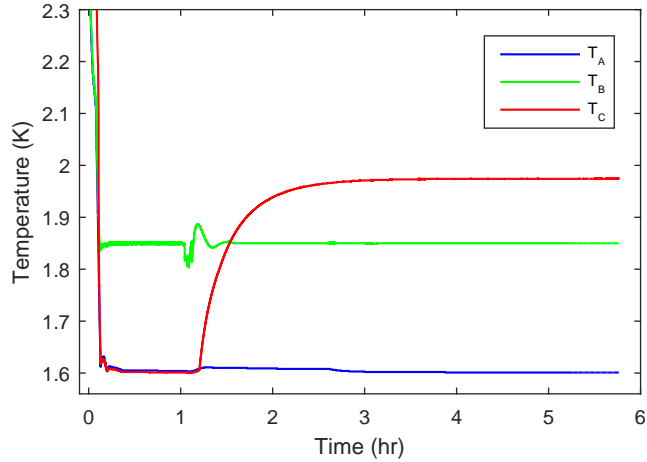}
\caption{A heating process of cell $C$.}
\label{fig:heating}%
\end{center}
\end{figure}

A heating process of cell $C$ is plotted in Fig. \ref{fig:heating}. At
 the initial moment represented
in the figure,  pot $A$ is filled
with liquid $^4$He at a temperature above $\lambda$ point while both pot $B$ and cell
 $C$ are
empty. Pumping pot $A$ leads to dropping of $T_A$. $T_B$ 
drops in pace with $T_A$
 due to relatively large thermal link between two pots. At some point,
 electric current of the resistance wire around pot $B$ is
activated  to stabilize $T_B$ 
at the set value of $1.85K$.
$T_C$ decreases with $T_A$. When  both of them 
are getting below $\lambda$ point,
superfluid  transport between  
pot $A$ and cell $C$ is initiated and consequently plays a major 
role in determining the value of $T_C$ relative to $T_A$. 
 $T_C$ remains a few millikelvins below $T_A$ for 
superfluid filling duration of cell $C$.  If $T_C$ is further below,
fountain pressure of the superfluid, 
caused by the difference between $T_A$ and $T_C$,
overcomes the gravitational pull and directs superfluid back from cell $C$ to
pot $A$, leading to increase of $T_C$. On the other hand, if $T_C$ is
getting closer to or further above $T_A$, the overall force (fountain 
pressure plus
the gravitational pull) conducts superflow from pot $A$ to cell
$C$, resulting in decrease of $T_C$. This negative 
feedback mechanism of temperature locks roughly the value of $T_C$ 
relative to $T_A$.

When cell $C$ is approximately fully filled and superfluid transport between cell
$C$ and pot $B$ is initiated, $T_B$ goes through some fluctuations.
 This is mainly due to the fact that heat capacitance of the empty
pot $B$ (made of copper) is relatively small. Even a small amount of 
cold superfluid  $^4$He entering pot $B$ can cause a large variation of
$T_B$ and the temperature stabilization system is not capable to respond 
in a timely manner.  
After a certain
amount of superfluid $^4$He is accumulated in pot $B$, further injection of 
cold superfluid is no longer capable to change 
  $T_B$ dramatically and the stabilization of  $T_B$ is well restored. 
 


Once the transport of superfluid $^4$He through superleak $S_{AC}$, cell $C$
and superleak $S_{BC}$ is fully established,  $T_C$ rises steadily and reaches a
value 120 $mK$ above  $T_B$.
The heat received by cell $C$ comes from superflows. Since the 
kinetic energies of superflows are negligible \cite{kinetic}, the heat must originate from 
the thermal energies of superflows. This heating phenomenon is some way 
analogous to the Peltier effect of an electric current flowing across two conductors:
the superflow entering cell $C$ carries a thermal 
energy (density) larger than that of the superflow leaving from cell $C$, leading 
to the heating of cell $C$. 

A natural understanding of this heating phenomenon can be provided 
by a recently developed  quantum theory of  superfluid $^4$He 
\cite{yu1,yu2, bloch, leggett, feynman}.  In quantum mechanics, 
microscopic processes of a liquid $^4$He system are governed by
the many-body Hamiltonian operator of the system and 
by the atomic-molecular interactions between the system
and its surroundings. One can use
the many-body eigen-states/levels of  the Hamiltonian operator and quantum
jumps among these eigen-states/levels (caused by the interactions 
between the system and its surroundings) to give    
 a full microscopic description of the system. When the temperature
is below the transition point, one can consider only 
the low-lying levels  since the high-lying
levels become irrelevant due to the Boltzmann exponential factor.
It is found that all the low-lying levels 
 fall into a large number of groups in a hidden way, with each level
belonging to one group only.  There exist high energy barriers which
separate different groups (of levels) and prevent inter-group quantum 
level transitions. On the other hand, the atomic-molecular interactions
between the system and its surroundings cause
 frequent intra-group level jumps 
in an occupied group, which leads to a thermal distribution in the group
 and to a group-specific thermal equilibrium between the system
and its surroundings. At a given temperature, the thermal 
distribution of levels in a group determines all (group-specific) 
macroscopic properties of the system, such as its thermal energy (density) and
 its flow velocity. Different groups have different flow velocities, thus 
one can use the flow velocities to distinguish groups and can further regard 
other group-specific properties as being flow-velocity-dependent.  
 It can be argued that the thermal energy density  has a 
negative dependence on flow velocity: 
 the larger the flow velocity is, the smaller the thermal energy density. This unusual
velocity dependence of thermal energy density  is responsible for
the fundamental coupling between the thermodynamic motion 
and the hydrodynamic motion of superfluid $^4$He , and it  
can be used to explain naturally the
mechano-caloric effect of the system.

In this experiment, the superflow velocities in superleaks behave in
 a rather subtle way. Note that a superflow is frictionless and its velocity 
can not be stabilized by friction, which is in contrast to the case of  an ordinary flow.    The 
superflow in superleak $S_{AC}$ keeps accelerating or decelerating, subject
 to pressure difference between superfluid in pot $A$ and that in cell $C$ (the fountain
pressure, caused by the temperature difference across $S_{AC}$, consists of
a significant part of the overall pressure difference).  
The pressure in   
cell $C$ rises rapidly when it changes from a nearly-fully-filled state to a fully-filled
 state, which regulates the inlet superflow (from the viewpoint of cell $C$) in a 
great deal and prevents it from getting a rather large velocity. On the other
 hand, the outlet superflow can be accelerated by the pressure rising in cell $C$ 
and can reach rather large velocity regime. Moreover, with $T_C> T_B$, the superflow 
in $S_{BC}$ can reverse flow direction when cell $C$ deviates 
from a fully-filled state, thus the actual flow time of outlet superflow is 
shorter than the flow time of inlet superflow. From these analyses it is clear  
that  there is an asymmetry 
between the velocity distribution of the inlet superflow and that of the outlet
 superflow,  which in turn leads to a difference between their thermal energies
  and  
to the heating of cell $C$.

In conclusion, we observed an intriguing heating effect of $^4$He superflows. 
This fundamental phenomenon 
can be explained naturally by a microscopic quantum theory 
of superfluid $^4$He.

This work was supported by the Chinese NSF (Grant No. 11474313), 
by CAS (Grant No. XDB21030300).
%




\begin{thebibliography}{10}
\bibitem{kinetic} The average transfer rate of superfluid helium from pot $A$ to pot $B$ is 3 to 4
cubic centimeters per hour, and the corresponding flow velocities in  superleaks are the order of
a few millimeter per second. The associated kinetic energy ( if applicable) is around $10^{-29}$ Joule
per $^4$He atom,  which corresponds to a temperature scale of 1 ${\mu}K$  (negligible 
for a helium system at a temperature above 1.6  $K$).   Also, as argued later in the manuscript,
the inlet superflow (from the viewpoint of cell $C$ ) has a velocity smaller than the outlet superflow,
so the change of kinetic energy (if applicable) of superflows shall lead to a tiny cooling of cell $C$
rather than heating it.
\end{thebibliography}
\end{document}